\documentclass[paper,11pt]{AAS}		

\usepackage{bm}
\usepackage{amsmath}
\usepackage{subfigure}
\usepackage[colorlinks=true, pdfstartview=FitV, linkcolor=black, citecolor= black, urlcolor= black]{hyperref}
\usepackage{overcite}
\usepackage{footnpag}			      	

\begin{document}

\title{The Taiwan Extragalactic Astronomical Data Center (TWEA-DC)}

\author{S\'ebastien Foucaud\thanks{\texttt{foucaud@ntnu.edu.tw}} 
\ and Yasuhiro Hashimoto \\
{\small {\it Department of Earth Sciences, National Taiwan Normal University, Taiwan}}}

\maketitle{}

\begin{abstract}
The next generation of telescopes and instruments are facilitating our understanding of the Universe by producing data at a pace that beats all projections, and astronomers today are left in the face of an avalanche of data like never before. In order to cope with this problem and come up with a reliable and innovative solution, Data Centers were created in various locations and the concept of Virtual Observatories elaborated. Based at the National Taiwan Normal University, the Taiwan Extragalactic Astronomical Data Center plan to join in global efforts by proposing 1Pb of data storage dedicated to extragalactic astronomy by 2015. In continuation with individual efforts in Taiwan over the past few years, this is the first stepping-stone towards the building of a National Virtual Observatory. \\
\indent Besides the common functionalities generally provided by data centers, our goal is to propose "on-the-fly" photometry measurements from publicly available surveys: a unique way for cross-matching information. Also we will propose access to raw and reducible data available from archives worldwide, a goldmine of under-exploited information. Finally, we will propose our own specific analysis tools available on-line through a user-friendly interface.\\
\indent Purchased very recently, the current Data Storage Unit is capable of accumulating up to 50Tb of data. In the first phase, we will focus on multiband catalog cross-matching and make the latest extragalactic datasets available to the worldwide community, which should be fully functional in 2011.
\end{abstract}

\section{Introduction}

Planning of the next generation of telescopes and instruments are becoming very ambitious requiring massive aggregation of resources and expertise. Project such as ALMA\footnote[2]{\texttt{http://www.almaobservatory.org}} and the Thirty-meter Telescope\footnote[3]{\texttt{http://www.tmt.org}}, will allow us to push forward our exploration to the edge of the Universe and help us survey the whole sky at a pace never imagined before. Furthermore, to gather the maximum possible information, these projects cover the sky in different wavebands, from gamma- and X-rays, optical, infrared, through to radio. Such observations require a wide range of expertise and information, which is sometimes split and difficult to bridge.
These breakthroughs in telescopes, detectors, and also computer technology allow astronomical instruments to produce several terabytes of images and catalogs. Astronomy, today, faces a data avalanche. It is already almost easier to Òdial-upÓ a part of the sky than wait many months to have access to a telescope. With the advent of inexpensive storage technologies and the availability of high-speed networks, the concept of multi-terabyte on-line databases interoperating seamlessly is no longer outlandish. More and more catalogs are now interlinked, crossing wavelengths boundaries. Furthermore the new generation of survey telescopes (Pan-STARRS, LSST, etc) will image the entire sky every few days and yield Petabytes of data. \\
\indent Over the past decade the concept of the Virtual Observatory (VO) has emerged rapidly to address challenges relating to data management, analysis, distribution and interoperability. The VO is a system in which the vast astronomical archives and databases around the world, together with analysis tools and computational services, are linked together into an integrated facility.
By providing the tools to assemble and explore massive data sets quickly, the VO facilitates and enables a broad range of sciences. Amalgamating massive data sets over a broad range of wavelengths, spatial scales, and temporal intervals is especially fruitful. VO-based studies include systematic explorations of the large-scale structure of the Universe, the structure of our Galaxy, AGN populations in the universe, variability on a range of time scales, wavelengths, and flux levels. The VO also enables searches for rare, unusual, or even completely new types of astrophysical objects and phenomena. For the first time, we are able to compare the results of massive numerical simulations with equally voluminous datasets. The International Virtual Observatory Alliance\footnote{\texttt{http://www.ivoa.net}} (IVOA) was formed in June 2002 with a mission to "facilitate the international coordination and collaboration necessary for the development and deployment of the tools, systems and organizational structures necessary to enable the international utilization of astronomical archives as an integrated and interoperating virtual observatory." The IVOA now comprises 15 national and three regional/agency VO projects. In East Asia, Japan and China are members of IVOA, who are also developing their own national VOs. 

\section{The Taiwan Extragalactic Astronomical Data Center}

\subsection{Philosophy}

Data centers contribute to global efforts in different ways: data archives, with a particular emphasis put on 'science ready' data; added-value databases, services; tools, software suites and algorithms, for instance for data visualization, data analysis and data mining; thematic services to help solve a well-defined science question; full data analysis or research environments. New types of services are emerging, in particular, theoretical services providing modeling results or matching models with observations. The main role of Data Centers is not only to provide a good quality service to the community, but also provide added value based on expertise. This requires shared efforts not only in developing software and database environments, but also in crossing information between observational projects of diverse nature and of different wavebands.\\ 
\indent The Taiwanese astronomical community needs to step into the VO era. Even though the results of the efforts made by the VO community worldwide are meant to be public, Taiwan must participate in it to prepare our next generation of astronomers who will require such skills and also not to be relegated to the ÔfollowersÕ position. To help us realize this vision for the future, NTNU has funded in 2010 the creation of the first Taiwan based Data Center dedicated to extragalactic astronomy. Several individual efforts have been conducted over the past few years to develop VO in Taiwan and the Taiwan Extragalactic Astronomical Data Center (TWEA-DC) is the obvious next leap forward. By having a fully functioning data center around which the community can work, Taiwan will be ready to join the international VO community. The efforts conducted by the VO community are already in very advanced stages, and therefore we will work on the base of their latest developments, and will include the available applications developed by the international community over the past decades. 

\subsection{Mission and Goals}

One of the major goals of the TWEA-DC is to form the next generation of astronomers, who will have to keep up pace with the changing face of modern Astronomy. Moving into the VO era will have a dramatic impact on the existing skill base of young astronomers. Therefore, by making a move now in this direction, Taiwan will prepare the next generation of scientists to face the technological revolution. Astronomy is now based on of large datasets, covering a broad wavelength range. The challenge is to aggregate the information and generate a final product that will bridge different expertise and, therefore, generate an enhanced scientific output. Therefore large amounts of data storage are required locally to enable a fast access to images and catalogs.  In order to fulfill this goal using gigantic amount of data, new tools for data analysis have to be developed. The TWEA-DC will help to fulfill this mission by focusing on three main goals:\\
\indent $\bullet$ {\it Matching different data-sets}: several million of objects cross-matched in very short timescale, requiring new algorithms and new concepts. Direct matching of  images will be actually ideal, and performing "on-the-fly" photometry and extraction is the way of the future algorithm. However as a first approach we plan to cross-match catalogs. Thanks to our algorithm, we are proposing a "on-the-fly" matching, enabling a new way of dealing with datasets.\\
\indent $\bullet$ {\it User-friendly portal to archived raw and reducible data}: standard data reduction is not always optimum, and dedicated processes are sometimes required. A centralized access to raw data will ease their exploitation.\\
\indent $\bullet$ {\it Specialized and dedicated analysis tools}: We are planning to develop and distribute new analysis tools through a user-friendly interface. Some of our tools are already ready for implementation, such as an online very fast correlation function measurements tool.

\section{Catalogs, images and "on-the-fly" matching tool}

\begin{figure}[htb]
	\centering\includegraphics[width=3in]{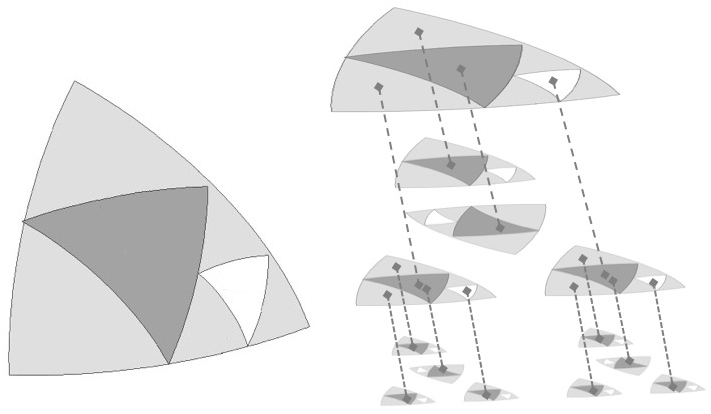}
	\caption{Hierarchical Triangular subdivision. The sky is divided in a succession of spherical triangles, which can be distributed in the form of a tree. The smallest elements will contain only one object. Such data structure enables very fast queries and cross-matches.}
	\label{fig:htm}
\end{figure}

As a initial service, the TWEA-DC provides a tool to cross-match multiband datasets to the community. Astronomers have to investigate innovative solutions to deal with the gigantic number of objects provided by new datasets. Services as simple as a data-base query or matching on the sky can be a real problem. One of the most recent solution is the use of hierarchical subdivision of the celestial sphere using spherical triangles. This kind of algorithms, based on quadtree algorithm, are nowadays widely used to query in astronomical database; for instance refer to the Hierarchical Triangular Mesh (HTM) algorithm\cite{2007cs........1164S} (see Fig.~\ref{fig:htm}). HTM is now a standard spatial indexing for astronomy and is used in various surveys (DES, LSST, etc). Based on this technology we developed a very fast code which allows us an "on-the-fly matching". This strategy enable the users to tune their own match, and to upload private catalogs and match them against public catalogs in the database.\\
\indent To really increase drastically the velocity of matching, new algorithm should not be based solely on the position on the sky, but make the best of the available wide range of parameters (fluxes, shapes, compactness, etc.). Our ultimate goal would be to directly use the images in order to perform the matching process. However the challenges to face (different nature of data, PSF, etc) imply a long-term development in close collaboration with computer scientists.

\section{A portal to raw and reducible data}

Usually data centers are focused in providing for fully reduced and calibrated data sets, processed through conventional data reduction pipelines. However for some specific studies, astronomers want to process the data using alternative data reduction algorithm. It is for instance the case to extract very low-surface brightness features from images (low-surface brightness galaxies, or tidal residuals around local galaxies, etc.). This requires a fully independent data reduction process, especially in the near-infrared for which the sky subtraction algorithm usually remove these features. Some data/archive centers propose access to raw and calibration data, but for a limited number of data-sets/telescopes. We propose to create a portal that will give access in one single interface to an exhaustive list of raw and reducible data produced by different facilities/telescopes. 

\section{Some high-level "on-the-fly" analysis tools}

One important mission of a Data Center is to provide the community for user-friendly tools to perform high-level analysis remotely. For instance the Centre de Donn\'ees astronomiques de Strasbourg - CDS\cite{2000A&AS..143....1G} - had developed a large number of web-based applications to access and visualize astronomical data-sets. Some basic tools such as plotting and database queries have to be part of the package proposed by the TWEA-DC, as well as access to some of the wide amount of tools already developed by the VO community (for instance TOPCAT\footnote{\texttt{http://www.star.bris.ac.uk/{\raise.17ex\hbox{$\scriptstyle\mathtt{\sim}$}}mbt/topcat/}}). However our primary goal is to develop our own set of dedicated analysis tools. Beyond the matching procedure, we want to provide the community with a service that will take fully advantage of our server. Indeed, the amount of data available impose the users to  conduct their analyses remotely.

\subsection{EAG: galaxy 2-point angular correlation in a blink}

The first application available will measure the angular 2-pt correlation function \cite{1980lssu.book.....P} for very large samples in a very small timescale. The correlation function quantifies the clustering of galaxies, and provides for vital information on the evolution of large-scale structure, on galaxy formation and also probes cosmological parameters. However performing such a study on very large samples may require long computing time. Given the size of the next generation of sky surveys, improved algorithms have to be developed. In practice, the 2-point angular correlation is simply measuring the excess of pairs in the data sample compared to a random distribution at different scales. EAG is based on a similar algorithm than our matching code, spreading the position in a quadtree structure, and double-walks are performed to count the pairs\cite{2004ASPC..314..249G}. We have developed a very fast code, coupled with a web-interface, that allows to determine the 2-point correlation function of several million of objects in a few minutes. EAG will be available to the community by the end of 2011.

\subsection{The next generation of tools}

We already are working on the next generation of tools that will be proposed as a service.\\
\indent To have a complete view of our Universe, one need to be able to measure how distant galaxies are, which can be achieved by measuring their redshift. However spectroscopic datasets are difficult to get for a complete sample of the sky and we have to rely on photometric methods to determine the redshift (SED $\chi^2$ minimization fitting, neuronal networks, etc). We are adapting some existing codes to entitle a photometric redshift determination "on-the-fly" remotely from our data-base. \\
\indent Galaxies are living in more-or-less dense structures that have a huge impact on their evolution. However this structures are not always straightforward to extract and require well thought algorithms.  We are developing a Group and cluster finders that will work "on-the-fly" remotely.\\
\indent Also we are collaborating closely with computer scientists that will apply algorithms developed for Data Mining \cite{2010IJMPD..19.1049B} to help the astronomical community to make the best out of the Data Center. 

\section{Raising interests and skills for Virtual observatories in Taiwan}

The most important role of the TWEA-DC for the next couple of years will be to prepare and train the current and next generation of astronomers to the future kind of astronomy. The scale of datasets and the complexity of the data themselves will require a new type of astronomers that will be familiar with computer science and able to collaborate fully with specialists on data mining.  This problem is known for decades (sometimes called the {\it Fourth paradigm}\cite{fourthpar} - the first three being observation, theory and simulations) and astronomers all around the globe have developed these skills and competences already (as testified by the International Virtual Observatory Alliance).\\
\indent The astronomical community in Taiwan is aware of the urge to join the global effort. However the rather small size of the community prevented so far individual efforts to be successful on the long term. The creation of a local Data Center will be a focus point for such efforts to be maintained. In parallel of software development, we will organize training activities dedicated to the astronomical community. We will organize workshops in collaboration with the Department of Computer Science and Engineering, for which we are also planning to invite foreign specialists on VO and Data mining in Astronomy. In a longer term we would like to host a IVOA Interoperability Meeting or an Astronomical Data Analysis Software and Systems (ADASS) conference in Taiwan. The current generation of students has to be prepare to tackle the new generation of datasets. As part of their training program we will propose courses in collaboration with our colleagues computer scientist, and we will organize Student Summer Schools in Taiwan, inviting international specialists. 

\section{The first year}

\begin{figure}[htb]
	\centering\includegraphics[width=3in]{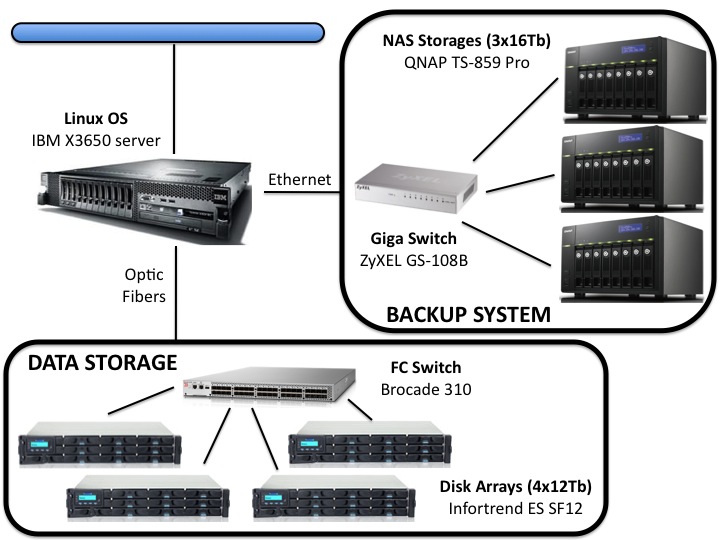}
	\caption{Current structure of the TWEA-DC.}
	\label{fig:DC}
\end{figure}

The current version of TWEA-DC gathers 48Tb of data storage. The Data center has been exclusively funded by the National Taiwan Normal University.  The structure of the TWEA-DC is tailored for rapid data access and is composed of a server, a data storage unit and a mirror backup System (see Fig.~\ref{fig:DC}). The communication speed between the server and the data units is of 4Gb/s, while the backup system operates at a rate of 1Gb/s. The Data Center is set up by a team of our graduate students from NTNU. They are responsible for implementing hardware, software, security, log, and backup system. They will also setup the database and create the web interface. The team works under our supervision with consultation of Computer scientists. We expect the Data Center to be ready for July 2011.The  Òon-the-flyÓ cross-matching between the major public multiband catalogs (SDSS, UKIDSS, CFHTLS, etc) and private catalogs matching will be the first service available. We expect to release our database and the associated tools for the community by the end of 2011.

\section{Conclusion}

The Taiwan Extragalactic Astronomical Data Center will be available to the worldwide community by the end of 2011. This effort, led by the National Taiwan Normal University, consists of the first stepping stone in building a National Virtual Observatory in Taiwan. Its main goals will be to make available the most important publicly available datasets along with a very fast catalog matching tool, allowing "on-the -fly" matching. The TWEA-DC will also provide a user-friendly portal to access easily reducible data from worldwide archives, and provide dedicated high end analysis tools. Finally the TWEA-DC will help to train the Taiwanese community, and will act as a bridge between the local astronomers, local computer scientists but also the global VO community.

\section{Acknowledgments}

The TWEA-DC is funded by the National Taiwan Normal University, and the Backup system by the NTNU-Department of Earth Sciences. SF would like to acknowledge the travel support for the EAMA conference in Shanghai from Academia Sinica, Institute of Astronomy and Astrophysics. We would like to thank SŽbastien Reybier (SoaMI), Nicolas Kamennoff (ACSEL) and the EAG team for their hard work in co-developing the matching algorithm and the software EAG. We also acknowledge the support from EPITECH and ACSEL during the development phase of EAG.

\end{document}